\documentclass[dvipsnames]{article}
\usepackage{graphicx}
\usepackage{booktabs}
\usepackage{xcolor}
\usepackage{caption}
\usepackage{subcaption}
\usepackage{showlabels}
\usepackage{enumitem}
\usepackage[ruled,vlined]{algorithm2e}
\PassOptionsToPackage{numbers, compress}{natbib}
\usepackage[final]{neurips_2020}

\usepackage[utf8]{inputenc} % allow utf-8 input
\usepackage[T1]{fontenc}    % use 8-bit T1 fonts
\usepackage{hyperref}       % hyperlinks
\usepackage{url}            % simple URL typesetting
\usepackage{booktabs}       % professional-quality tables
\usepackage{amsfonts}       % blackboard math symbols
\usepackage{nicefrac}       % compact symbols for 1/2, etc.
\usepackage{microtype}      % microtypography
\usepackage{hyperref}
\usepackage{multirow, makecell}

\definecolor{darkgreen}{rgb}{0,0.6,0}

\SetKwFunction{Range}{range}
\SetKwInput{KwInput}{Input}
\SetKwInput{KwParams}{Parameters}

\title{Pushing the Limits of Semi-Supervised Learning for Automatic Speech Recognition}
\author{%
  Yu Zhang\thanks{Equal contribution.} \And
  James Qin$^*$ \And
  Daniel S. Park$^*$ \And
  Wei Han \And
  Chung-Cheng Chiu \And
  Ruoming Pang \And
  Quoc V. Le \And
  Yonghui Wu \AND
  \\
  Google Research, Brain Team \\[7pt]
  \texttt{\{ngyuzh, jamesqin, danielspark, weihan, chungchengc, rpang, qvl, yonghui\}} \\
  \texttt{@google.com}
}

\begin{document}

\maketitle

\begin{abstract}
We employ a combination of recent developments in semi-supervised learning for automatic speech recognition to obtain state-of-the-art results on LibriSpeech utilizing the unlabeled audio of the Libri-Light dataset. More precisely, we carry out noisy student training with SpecAugment using giant Conformer models pre-trained using wav2vec 2.0 pre-training. By doing so, we are able to achieve word-error-rates (WERs) 1.4\%/2.6\% on the LibriSpeech test/test-other sets against the current state-of-the-art WERs 1.7\%/3.3\%.
\end{abstract}

\section{Introduction}

Recently, semi-supervised learning (SSL) methods have been used to drastically improve the performance of automatic speech recognition (ASR) networks. The goal of semi-supervised learning is to use a large unlabeled dataset to help with improving the performance of a supervised task defined by a labeled dataset. In this work, we combine recently proposed pre-training and self-training methods to obtain state-of-the-art (SOTA) performance on LibriSpeech \cite{librispeech}. We use the audio from the Libri-Light dataset \cite{librilight}, derived from the LibriVox database of free public domain audio books, as the unlabeled data for semi-supervised learning.

Recall that in iterative self-training, a series of models are trained where a given model in the series serves as a teacher to the succeeding model by generating labels on the unlabeled dataset. The student to this teacher model is trained on the dataset obtained by combining the supervised set with the teacher-labeled dataset. Meanwhile, in pre-training, the model (or a part thereof) is trained on the unlabeled data via a pre-training task, and then fine-tuned on the supervised dataset. 

Our approach is to combine iterative self-training and pre-training in a straightforward fashion. We pre-train a series of models, which are then used to initialize models for iterative self-training. Here, the unlabeled dataset serves a dual purpose of being used as a pre-training dataset, and also as the unlabeled dataset for which pseudo-labels for training student models are generated. This idea has been explored recently for image tasks in \cite{simclrv2, zoph2020rethinking}.

Some important components in our recipe for combined SSL are as follows:
\begin{itemize}
\item \textbf{Model:} We use Conformers \cite{conformer} (and slight variants thereof) as the model architecture.
\item \textbf{Pre-training:} We use wav2vec 2.0 pre-training \cite{wav2vec2} to pre-train the encoders of the networks.
%Instead of CTC, we use RNN-T model.
\item \textbf{Iterative Self-Training:} We use noisy student training (NST) \cite{nst, nstasr} with adaptive SpecAugment \cite{specaugment,specaugment2} as our iterative self-training pipeline.
\end{itemize}

By carrying out combined SSL using giant Conformers, we are able to obtain state-of-the-art performance on the LibriSpeech dev and test sets, achieving the word-error-rates 1.3\%/2.6\%/1.4\%/2.6\% on the dev/dev-other/test/test-other sets.\footnote{We emphasize that making the ASR model bigger on its own does not result in performance gains---only upon applying SSL methods do we reap the benefit of the enlarged model size, as demonstrated in section \ref{ss:model size}.}

\begin{figure}[h!]
\centering
\includegraphics[width=0.95\columnwidth]{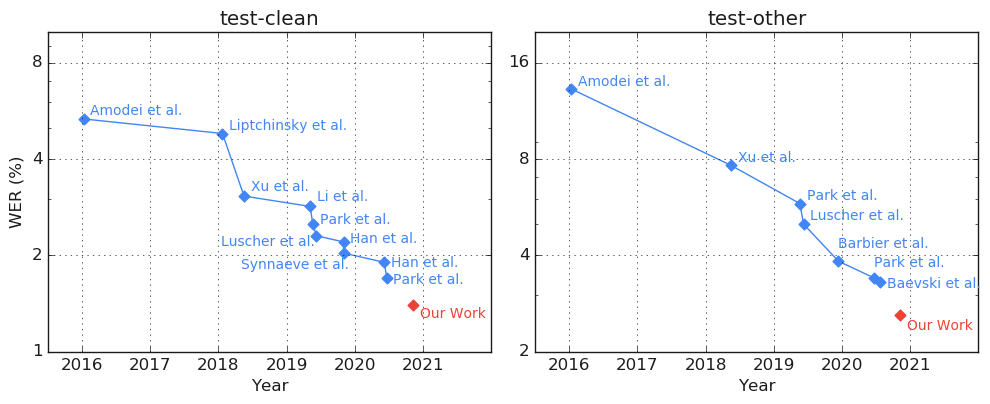}
\vskip 0.05in
\caption{History of state-of-the-art LibriSpeech test/test-other WERs (\%) according to Papers With Code. The y-axis is log-scaled.}
\label{f:history}
\end{figure}

In figure \ref{f:history}, we have plotted the history of LibriSpeech test and test-other performances reported according to Papers With Code \cite{testclean, testother} to put our work into context. The papers corresponding to each data point have been collected in the bibliography \cite{wav2vec2, nstasr, specaugment, amodei2016deep, liptchinsky2017letter, xu2018neural, li2019jasper, luscher2019rwth, han2019state, synnaeve2019end, contextnet, barbier20190}.

\subsection{Related Work}

There have been two major directions of research regarding semi-supervised learning. The first is self-training, where a teacher model is used to improve the performance of a student model by generating labels for the unlabeled set the student can train on \cite{scudder1965probability,yarowsky1995unsupervised,riloff2003learning}. Self-training has been a popular technique in ASR studied extensively in the literature \cite{Zavaliagkos98utilizinguntranscribed,Lamel00lightlysupervised,Novotney2009,Thomas2013,li2019,kahn2019selftraining,synnaeve2019endtoend,parthasarathi2019,hsu2020selfsupervised}.

Another big theme in semi-supervised learning is consistency \cite{xie2019unsupervised,zhai2019s4l,sohn2020fixmatch}. In this line of research, a consistency-based task that can be trained on unlabeled data is introduced and used to pre-train networks so that they can learn a good representation of the data. These networks are in turn fine-tuned on supervised data \cite{hsu2018extracting, chung2018speech2vec, chung2019autoregressive, chorowski2019unsupervised, schneider2019wav2vec, baevski2019vqwav2vec, ling2019deep}.

Our work aims to combine the benefits of both approaches by employing a series of pre-trained models in a self-training loop. The self-training method of choice is noisy student training \cite{nstasr}, that capitalizes on the application of augmentation, motivated by research in the image domain \cite{nst,he2020revisiting}. We pre-train our models using wav2vec 2.0 pre-training \cite{wav2vec2}, a recent method of pre-training influenced by progress in NLP research \cite{bert}.

\section{Methods}

We summarize important components of our pipeline in this section. These elements are utilized to conduct experiments on the LibriSpeech + Libri-Light task, as elaborated in the next section.

\subsection{Model Architecture: Conformer}

\begin{figure}[h!]
\centering
\includegraphics[width=0.8\columnwidth]{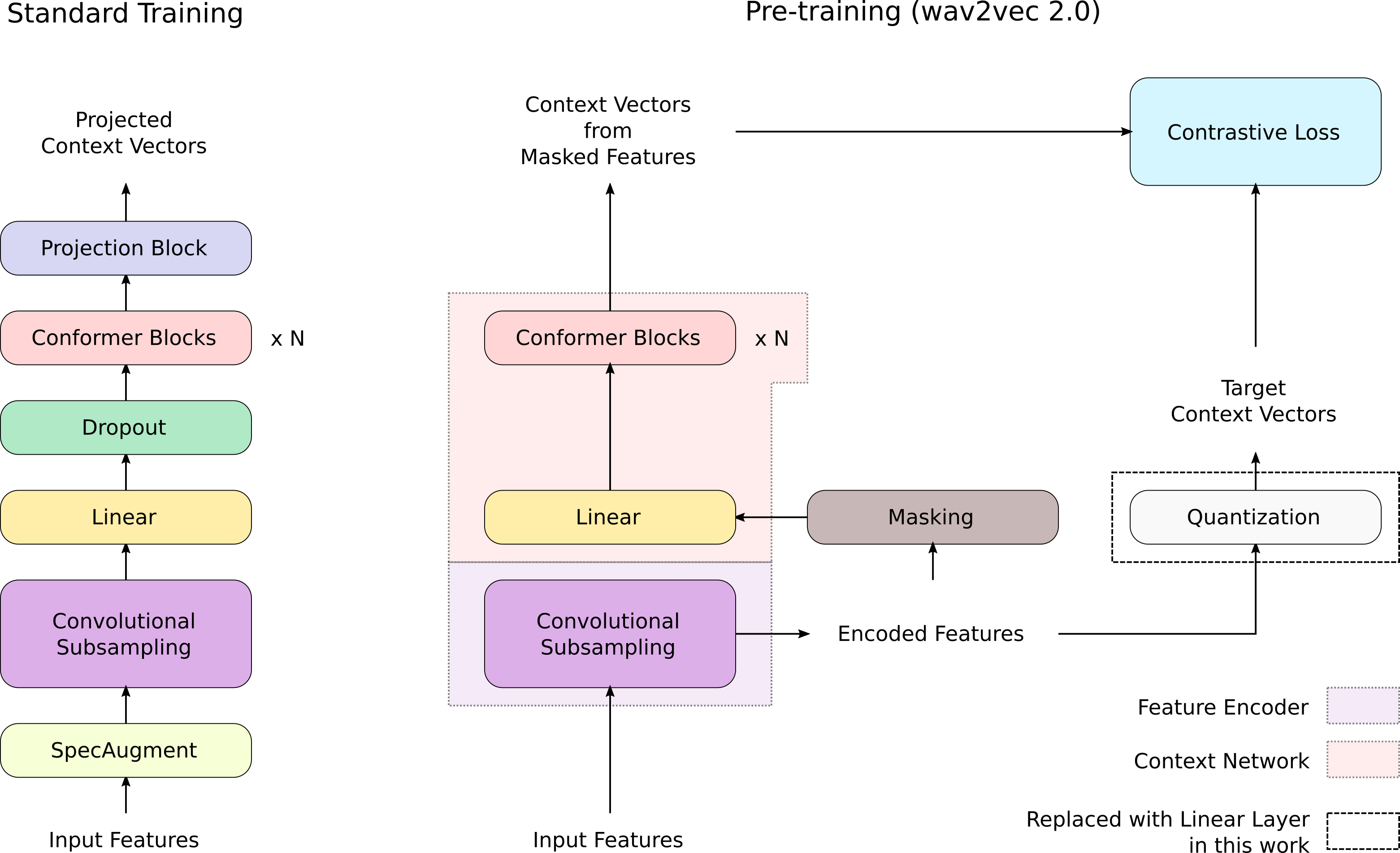}
\vskip 0.05in
\caption{The Conformer encoder. In wav2vec 2.0 pre-training, the features generated by the convolutional sampling block are masked and passed into the rest of the network to yield context vectors, and also quantized to yield target context vectors. The contrastive loss between the context vectors obtained from masked features and the quantization unit is optimized. In our work, we replace the quantization layer with a linear layer. During fine-tuning, an additional projection block is added to produce features to be passed to the transducer.}
\label{f:conformer}
\end{figure}

The ASR network is a sequence transducer \cite{rnnt} consisting of a LSTM \cite{lstm} decoder and a Conformer \cite{conformer} encoder, whose major component is a stack of "conformer block"s, each of which is a series of multi-headed self attention \cite{vaswani2017attention}, depth-wise convolution and feed-forward layers. For large models used in this work, we get rid of relative positional embedding \cite{dai2019transformer} from the self attention layer, which greatly speeds up training without impairing the model quality. The overall structure of the Conformer encoder is depicted on the left-hand-side of figure \ref{f:conformer}.

We introduce scaled-up versions of the Conformer which we denote Conformer XL and Conformer XXL, which have 600M and 1B parameters, respectively. Details on the architecture are listed in table \ref{t:conformers} in comparison to "Conformer L" used in the original work \cite{conformer}. Conformer L has a single-layer LSTM as its decoder, while XL and XXL have two-layer LSTM decoders. All decoders have dimension 640. We use a linear layer with Swish activation \cite{ramachandran2017searching} and batch-normalization as the projection block for these models.

\begin{table}[h!]
  \caption{Parameters for Conformer models.}
  \vskip 0.1in
  \label{t:conformers}
  \centering
  \resizebox{\columnwidth}{!}{%
  \begin{tabular}{lccccccc}
    \toprule
    Model & \# Params (B) & Enc. Layers & Enc. Dim & Att. Heads & Conv Kernel Size & Rel. Att. & \# dec. layers\\
    \midrule
    Conformer L & 0.1 & 17 & 512 & 8 & 32 & Y & 1\\
    Conformer XL & 0.6 & 24 & 1024 & 8 & 5 & N & 2\\
    Conformer XXL & 1.0 & 42 & 1024 & 8 & 5 & N & 2\\
    \bottomrule
  \end{tabular}}
\end{table}

We also introduce a model we denote Conformer XXL+, which is obtained by adding an additional Conformer block and a stacking layer to the Conformer XXL. The stacking layer reshapes the input to have half the time length and twice the channel size. As explained later, it will be convenient to view this model as a Conformer XXL model with a bigger projection block. Conformer XXL+ has approximately 50M more parameters compared to Conformer XXL.

\subsection{Wav2vec 2.0 Pre-training}

We pre-train the Conformer encoder akin to wav2vec 2.0 pre-training \cite{wav2vec2} with 60k hours of unlabeled audio from the "unlab-60k" subset of Libri-Light \cite{librilight}. Unlike in the original work which takes raw waveforms as input, we use log-mel spectrograms. The Conformer encoder could naturally be split into a "feature encoder" consisting of the convolution subsampling block and a "context network" made of a linear layer and a stack of Conformer blocks. The convolutional subsampling block has two 2D-convolution layers, both with strides $(2,2)$, resulting in a 4x reduction in the feature sequence length. The encoded features from the convolutional subsampling block, on the one hand, are fed into the rest of the network after masking to yield context vectors, and on the other, are passed through a linear layer (rather than a quantization layer as in the original work) to produce target context vectors. Wav2vec 2.0 pre-training optimizes the contrastive loss between the context vectors from the masked positions and the target context vectors. This procedure is depicted on the right-hand-side of figure \ref{f:conformer}.

We note that a projection block that produces features passed to the transducer from the context vectors, is added and trained at the fine-tuning stage. We assume this block can have varying degrees of complexity. Taking this point of view, we consider the Conformer XXL and Conformer XXL+ as having the same pre-trained sub-model (feature encoder and context network) with different projection blocks. While the Conformer XXL has a single linear layer (with Swish activation and batch-normalization) as the projection block, the projection block of Conformer XXL+ is composed of a conformer block and a stacking layer followed by the aforementioned linear layer. Thus the Conformer XXL and XXL+ models are initialized with the same pre-trained sub-model at the fine-tuning stage.

\subsection{Noisy Student Training with SpecAugment} \label{ss:nst}

We use the noisy student training pipeline for ASR studied in \cite{nstasr} for training the models pre-trained with wav2vec 2.0. In NST for ASR, a teacher model, obtained by shallow-fusing \cite{shallowfusion} an ASR model with a language model (LM), is used to generate transcripts for the unlabeled data via inference on un-augmented audio. The teacher-labeled data, after filtering and balancing, are then used to train the next generation ASR model. The input data to the student model is augmented using adaptive SpecAugment \cite{specaugment,specaugment2}. For obtaining the main result, we find that it is most effective to use the entirety of the teacher-generated data, rather than filtering and balancing them.

To summarize, with the labeled LibriSpeech dataset $S$, the unlabeled Libri-Light dataset $U$ and an LM trained on the LibriSpeech LM corpus, the following procedure is used to train a series of models:
\begin{enumerate}
\item Fine-tune pre-trained model $M_0$ on $S$ with SpecAugment. Set $M = M_0$.
\item Fuse $M$ with LM and measure performance.
\item Generate labeled dataset $M(U)$ with fused model.
\item Mix dataset $M(U)$ and $S$.
\item Fine-tune new pre-trained model $M'$ with SpecAugment on mixed dataset.
\item Set $M = M'$ and go to 2.
\end{enumerate}

\section{Experiments}

We use the NST pipeline to train a series of pre-trained Conformers to obtain SOTA performance on LibriSpeech dev and test sets by utilizing unlabeled audio from the Libri-Light dataset.

\subsection{Experiment Settings}

\textbf{Data:} We aim to improve the LibriSpeech task \cite{librispeech} by utilizing the unlabeled audio in the "unlab-60k" subset of the Libri-Light \cite{librilight} dataset. The 960h of transcribed audio of the LibriSpeech dataset is used as the supervised data. We use 80-dimensional log-mel filter bank coefficients of the utterances as single-channel input features for the networks. A 1024-token word-piece-model (WPM) \cite{wpm} constructed from the transcripts of the LibriSpeech training set is used to tokenize the transcripts.

We process the Libri-Light dataset slightly differently for pre-training and fine-tuning within the NST loop. The audio dataset for wav2vec 2.0 pre-training is obtained by randomly segmenting the audio from the Libri-Light dataset to have segmentation length between 32 and 64 seconds, resulting in approximately 2.9 million utterances in total 30,031 hours of data. We further randomly sample 32 seconds chunks from this audio on-the-fly during pre-training, as we found it to lead to more efficient training without degradation of performance. For fine-tuning, we follow the same data preprocessing procedure as in \cite{nstasr}.  
\smallskip

\textbf{ASR Models:} We train four generations of models, including the zeroth generation only trained on the labeled LibriSpeech dataset. We use the pre-trained Conformer XL at generations 0 and 1, and use the pre-trained Conformer XXL at generations 2. At generation 3, we experiment with both the pre-trained Conformer XXL and XXL+. As explained in the previous section, both models are initialized with the same pre-trained portion of Conformer XXL.
\smallskip

\textbf{SpecAugment:} We use an adaptive SpecAugment \cite{specaugment, specaugment2} policy with two frequency masks with mask size parameter $F = 27$, and ten time masks with maximum time-mask ratio $p_S = 0.05$ to augment the input for the student model in each generation. We use these same parameters for all generations.
\smallskip

\textbf{Batch-wise Mixing:} For all generations, the ratio of supervised versus teacher-labeled data in each batch is fixed to 1:9 during training.
\smallskip

\textbf{Language Model and Fusion:} An eight-layer 103M-parameter transformer language model \cite{vaswani2017attention} with relative positional embedding \cite{dai2019transformer} has been trained on the LibriSpeech language model corpus, along with the transcripts from the 960h training set of LibriSpeech. The model achieves perplexity-per-word 59.44 on the LibriSpeech dev set transcripts. Following \cite{nstasr}, we re-tune the LM fusion parameters (the LM fusion weight \cite{shallowfusion} and the non-blank reward \cite{sak2015fast,zhang2020transformer}) at each generation to maximize dev set performance of the fused model. 
\smallskip

\textbf{Pre-training Parameters:} We use identical masking parameters as \cite{wav2vec2} to mask the features generated from the convolutional subsampling block, i.e., we sample the initial time step of the mask randomly with probability 0.065 and mask the subsequent 10 steps. We train with global batch size 2048 on 256/512 Google TPU V3 cores for 3-4 days for the XL/XXL models respectively. For the XL model, we use Adam optimization with a transformer learning rate schedule (section 5.3 of \cite{vaswani2017attention}) with peak learning rate 2e-3 and 25k warm-up steps. We apply gradient scaling to cap the norm of the gradient to 20. For the XXL model we switch to Adafactor \cite{adafactor} with parameters $\beta_1=0.9$ and $\beta_2=0.98$, and use 2nd-moment estimator factorization to reduce the accelerator memory footprint. The learning rate schedule stays the same.
\smallskip

\textbf{Fine-tuning Parameters:} We fine-tune the pre-trained checkpoints (400k steps) with global batch size 1024/512 on 256/512 Google TPU v3 cores for 1-3 days for the XL/XXL models in the NST loop. We optimize the encoder and decoder with separate optimizers and learning rate schedules given that the encoder alone has been pre-trained. For the XL model, we use Adam optimization with a transformer learning rate schedule (section 5.3 of \cite{vaswani2017attention}). For the encoder, we use peak learning rate 3e-4 with 5k warm-up steps, while for the decoder, a peak learning rate 1e-3 and 1.5k warm-up steps are used. For the XXL models, the Adafactor optimizer with the same configuration as in pre-training is used, with the same learning rate schedule as the XL model. For evaluation, we keep a separate copy of exponential-moving-averaged model weights aggregated with decay rate 0.9999.

\smallskip

\subsection{Experiment Results}

\begin{table}[h!]
  \vskip -0.05in
  \caption{WERs(\%) from LibriSpeech experiments. We compare models trained without any unlabeled data (baseline), trained using noisy student training without any pre-training (NST only), fine-tuned from a pre-trained model only using supervised data (pre-training only) and the model obtained by our combined SSL pipeline.}
  \vskip 0.1in
  \label{t:librispeech}
  \centering
  \small
  \resizebox{0.95\columnwidth}{!}{%
  \begin{tabular}{lccccccccc}
    \toprule
    \bfseries Method & \multirowcell{2}{\\[-7pt]Unlabeled data\\(hrs)}
    & \multicolumn{4}{c}{\bfseries No LM} & \multicolumn{4}{c}{\bfseries With LM} \\
    \cmidrule(r){3-6} \cmidrule(r){7-10}
    & & \bfseries dev & \bfseries dev-other & \bfseries test & \bfseries test-other
     & \bfseries dev & \bfseries dev-other & \bfseries test & \bfseries test-other \\
    \midrule
    \bfseries Baseline \\
    \quad  Conformer L \cite{conformer}
    & -
    & 1.9 & 4.4 & 2.1 & 4.3 
    &  &  & 1.9 & 3.9 \\
    % \quad Conformer XL & 2.1 & 4.9 & 2.3 & 4.9 \\
    \midrule
    \bfseries NST Only \\
    \quad Gen4 ContextNet \cite{nstasr}
    & 60k
    & 1.6 & 3.7 & 1.7 & 3.7
    & 1.6 & 3.4 & 1.7 & 3.4 \\
    \quad Gen4 Conformer L
    & 60k
    & 1.6 & 3.3 & 1.7 & 3.5
    & 1.6 & 3.1 & 1.7 & 3.3 \\
    \midrule
    \bfseries Pre-training Only \\
    \quad Pre-trained CTC \cite{wav2vec2}
    & 60k
    & 2.1 & 4.5 & 2.2 & 4.5 & 1.6 & 3.0 & 1.8 & 3.3 \\ 
    \quad Pre-trained Conformer XL
    & 60k
    & 1.7 & 3.5 & 1.7 & 3.5 
    & 1.6 & 3.2 & 1.5 & 3.2 \\
    \quad Pre-trained Conformer XXL
    & 60k
    & 1.6 & 3.2 & 1.6 & 3.3 
    & 1.5 & 3.0 & 1.5 & 3.1 \\
    \midrule
    \bfseries Combined SSL \\
    \quad  Gen3 Conformer XXL
    & 60k
    &  \bfseries 1.3 & 2.7  & 1.5  & 2.8 
    &  \bfseries 1.3 & \bfseries 2.6  & \bfseries 1.4  & 2.7  \\
    \quad  Gen3 Conformer XXL+
    & 60k
    &  \bfseries 1.3 & 2.7  & 1.5  & 2.7 
    &  \bfseries 1.3 & \bfseries 2.6 & \bfseries 1.4 & \bfseries 2.6 \\
    \bottomrule
  \end{tabular}
  }
\end{table}

The result of training Conformers with our pipeline is presented in table \ref{t:librispeech}, against some baselines. As we present in the next section, we have been unsuccessful in achieving gains from training Conformer XL and XXL from scratch, and only present the performance we obtained by fine tuning pre-trained models on LibriSpeech for these models.

As one of the baselines, we train a Conformer L model with NST without pre-training, using ContextNets \cite{contextnet} as intermediate models. We follow the exact procedure of \cite{nstasr}, in which five generations of ContextNets are trained, up to the fourth generation, then replace the fifth generation model with the Conformer L.

\begin{figure}[h!]
\centering
\includegraphics[width=0.8\columnwidth]{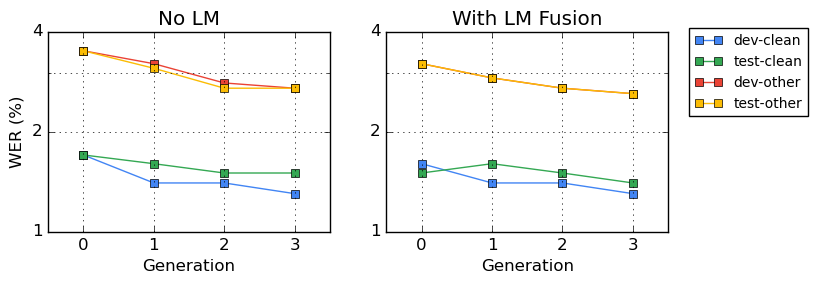}
\vskip 0.05in
\caption{Performance of models on dev/test sets of LibriSpeech at each generation with and without LM fusion. The dev and test-other performances of LM-fused models coincide at the level of precision considered. The final generation model is taken to be Conformer XXL+. The y-axis is log-scaled.}
\label{f:gens}
\end{figure}

We find that the generation-3 Conformer XXL model shows a 7-15\% relative improvement in WER across the dev and test sets compared to the pre-trained baseline. The performance of models at each NST generation, with Conformer XXL+ as the last generation model, are plotted in figure \ref{f:gens}.

\section{Discussion}

\subsection{Model Size and Pre-training} \label{ss:model size}

Here we present a set of experiments where we train models of various sizes and observe the effect of pre-training. To make fair comparisons, we experiment with a modified version of Conformer L, where relative positional embedding has been removed.

As shown in Table \ref{t:scale_models}, we find that merely scaling up the model size from 100M to 1B parameters alone does not improve performance, as we found it difficult to get gains from training the larger models on the supervised dataset. Upon pre-training, however, we observe consistent improvement by increasing the model size up to 1 billion parameters. We see that pre-training enables the model size growth to transfer to model performance.
%In the meanwhile, scaling up model can also boost the performance for self-supervised learning.
\begin{table}[h!]
  \vskip -0.05in
  \caption{WERs(\%) from LibriSpeech experiments. LM fusion has not been used.}
  \vskip 0.1in
  \label{t:scale_models}
  \centering
  \small
  \resizebox{0.9\columnwidth}{!}{%
  \begin{tabular}{lccccc}
    \toprule
    \bfseries Method & \# Params (B) & \bfseries dev & \bfseries dev-other & \bfseries test & \bfseries test-other \\
    \midrule
    \bfseries Trained from scratch \\
    \quad Conformer L (no rel. attn.)  & 0.1 & 2.0 & 4.7 & 2.2 & 4.8 \\
    \quad Conformer XL & 0.6 & 2.1 & 4.9 & 2.3 & 4.9 \\
    \quad Conformer XLL & 1.0 & 2.3 & 5.5 & 2.6 & 5.6 \\
    \midrule
    \bfseries With pre-training \\
    \quad Pre-trained Conformer L (no rel. attn.) & 0.1 & 2.0 & 4.6 & 2.0 & 4.5 \\
    \quad Pre-trained Conformer XL & 0.6 & 1.7 & 3.5 & 1.7 & 3.5 \\
    \quad Pre-trained Conformer XXL & 1.0 & 1.6 & 3.2 & 1.6 & 3.3 \\
    \bottomrule
  \end{tabular}
  }
\end{table}

\subsection{Ablations for Pre-training}

As explained previously, our pre-training procedure differs from that of \cite{wav2vec2} in a number of ways. First, we use the log-mel spectrogram instead of the waveform as the input for the ASR network. Secondly, we do not use quantization or a diversity loss for simplicity. It would be interesting to further investigate quantization and diversity loss settings that improve the performance of Conformers.

To reduce the computation and memory cost for the larger models, the length of the time series of the input features are reduced by the convolutional subsampling block. In table \ref{t:pretraining}, we present multiple experiments with different audio segment lengths for training with multiple time reduction factors by varying the strides of the convolutional layers in the convolutional subsampling block. We also experiment with using voice activation detection (VAD) with tools provided with \cite{librilight} to segment the inputs. For quick experimentation, we fine-tune the pre-trained models on the 100-hour clean subset of LibriSpeech.

We find that the convolutional subsampling layer is required to reduce the input length more aggressively when using longer input segments to pre-train the model. We also find that using VAD tends to degrade the performance, as we cannot guarantee that all the segments have enough context to benefit pre-training.

\begin{table}[h!]
  \vskip -0.05in
  \caption{WERs(\%) from fine-tuning pre-trained Conformer XL on LibriSpeech 100h. LM fusion has not been used.}
  \vskip 0.1in
  \label{t:pretraining}
  \centering
  \resizebox{0.8\columnwidth}{!}
  {%
  \begin{tabular}{ccccccc}
    \toprule
    VAD & Segment length (s) & Reduction & \bfseries dev & \bfseries dev-other & \bfseries test & \bfseries test-other \\
    \midrule
    No & 16 & 4x &  3.4 & 7.2 & 3.4 & 7.0 \\
    No & 16 & 2x & \bfseries 2.5  & 5.5 & 2.6 & 5.6 \\
    No & 32 & 4x & \bfseries 2.5  & \bfseries 4.7 & 2.6 & 4.9 \\
    Yes & 0 to 36 & 4x & 2.8 & 5.6 & 2.8 & 5.7 \\
    \bottomrule
  \end{tabular}
  }
\end{table}

\subsection{Ablations for Fine-tuning}

We present the results of ablation studies on the NST parameters in table \ref{t:finetuning}. These studies are conducted at generation 1 of the noisy student training pipeline, where a Conformer XL model is trained with transcripts generated by the fused generation-0 Conformer XL teacher-model.

\begin{table}[h!]
  \vskip -0.05in
  \caption{WERs(\%) from the generation-1 Conformer XL with different mix ratios and data processing policies. LM fusion has not been used.}
  \vskip 0.1in
  \label{t:finetuning}
  \centering
  \small
  \resizebox{0.9\columnwidth}{!}{%
  \begin{tabular}{lcccccc}
    \toprule
    & B.W. Mix Ratio & Data Processing & \bfseries dev & \bfseries dev-other & \bfseries test & \bfseries test-other \\
    \midrule
    \bfseries Selected & 1:9 & None & \bfseries 1.4 & 3.2 & 1.6 & 3.1 \\
    \midrule
    \bfseries Mix Ratio & 2:8 & None & 1.5 & 3.2 & 1.6 & 3.1 \\
    & Not Used & None & 1.5 & \bfseries 3.1 & 1.6 & 3.0 \\
    \midrule
    \bfseries Data Processing & 1:9 & Balanced & 1.5 & 3.2 & 1.5 & 3.1 \\
    & 1:9 & LM-filtered & 1.5 & 3.2 & 1.6 & 3.2 \\
    \bottomrule
  \end{tabular}
  }
\end{table}

We experiment with three different batch-wise mixing settings, where the ratio of the supervised versus teacher-labeled utterances were mixed with ratio 2:8, 1:9 or not mixed batch-wise, but randomly distributed. We find the 1:9 mix ratio and non-batch-wise mixing to be comparable, while the 2:8 mix ratio gives the worst performance.

In \cite{nstasr}, filtering teacher-generated transcripts using the confidence derived from the fused ASR model was shown not to be beneficial for the NST pipeline for the LibriSpeech + Libri-Light task. On the other hand, the authors found balancing the teacher-generated transcripts using sub-modular sampling useful. We have thus experimented with balancing, and filtering based on the log-perplexity score of generated transcripts computed by the LM trained on the LibriSpeech LM corpus, rather than the confidence of the ASR model.

The balancing procedure of \cite{nstasr} uses batch-wise optimization to assign sampling weights to the teacher-generated transcripts so that the KL divergence between the token distribution of the transcripts of the training set and the weighted teacher-generated transcripts is minimized. We use the exact same procedure as in \cite{nstasr} with equivalent parameters to balance the teacher-generated transcripts and train the generation-1 model for comparison.

Meanwhile, we also experiment with LM-filtering, motivated by the fact that the LM can capture more subtle features of the distribution of the LM corpus beyond token frequency. We compute the token length and LM score of all the transcripts generated by the fused generation-0 Conformer XL model, and compute the normalized filtering score of the transcripts using equation (1) of \cite{nstasr}. We use this score to filter out 40\% of the transcripts and use the remaining transcripts to train the model.

For this particular task and series of models, it turns out that neither balancing nor LM-filtering helps with generation-1 model dev performance. We hypothesize that this might be due to the fact that the generation-0 model is already extremely good, and that maximizing the amount of data produced by the model should be prioritized over improving the quality by a marginal amount.

\section{Conclusion}

We have combined recent developments in architecture, augmentation and especially semi-supervised learning to push the state-of-the-art on the speech recognition task LibriSpeech. 

\begin{ack}
We thank Françoise Beaufays, Yuan Cao, Shefali Garg, Parisa Haghani, Dongseong Hwang, Ye Jia, Bo Li, Manasa Prasad, Khe Chai Sim and Trevor Strohman for useful discussions. We thank Zhifeng Chen for helping us with scaling up the size of the models used in this work.
\end{ack}

\small
\bibliography{references}
\bibliographystyle{unsrtnat}

\end{document}